\documentstyle[aps,12pt,fleqn]{revtex}

\begin{document}
\title{Novel type of color solitons with topological charge \\
not coinciding with baryon number and \\
extended chiral group E$\chi $ }
\author{Victor Novozhilov and Yuri Novozhilov \\
V.A.Fock Institute of Physics, St.Petersburg State University,
198504 St.Petersburg, Russia  \\
E-mail: yunovo@pobox.spbu.ru, vnovozhilov@mail.ru}

\maketitle

\begin{abstract}
We consider extended chiral group $E\chi $ of gauge plus chiral
transformations of the quark path integral in the background gauge and
present novel type of color solitons described by the effective action
resulting from bosonization. $O(3)$ solitons are formed by a chiral field 
including parameters with diquark quantum numbers within symplectic 
subgroup of $E\chi $ , as well as within complete $E\chi $; isolated solitons
(i.e. solitons in gluonic vacuum) can be stable due to proper asymptotics 
induced by background vacuum, their topological charge starts from 
$\pm \frac 43$ and does not coincide with the baryon number, as if notions 
of left anf right were dependent also on non-abelian directions in such a 
$SU(2)$ group, where the baryon number defines a diagonal generator.
\end{abstract}

\pacs{12.39.Fe, 11.30.Rd, 11.10.Lm}

Understanding world of color solitons is important in an non-perturbative
approach to the low energy hadron physics. One of the problems is to include
color chiral anomalies \cite{Adler+Bell+Jackiw} in the total (gauge plus
chiral,or anomalous) color space and to study changes of gauge space.

In the gauge sector of $SU(2)$ QCD theory, color solitons were recently
reported by Faddeev-Niemi \cite{Faddeev+Niemi+Battye+Sutcliffe} and Cho et
al. \cite{Cho}. In the quark chiral sector, the Skyrmion model for
constituent quarks (qualitons) was discussed by Kaplan \cite{Kaplan}, and it
was shown recently \cite{Novozhilov-01} that isolated color solitons (i.e.
solitons on the background of vacuum gluon field) can be classically
stabilized by the chromomagnetic vacuum field in the cases of two colors,
one flavor and three colors, one flavor, their mass can be evaluated and
intersoliton potential displays confinement behavior.

The aim of this Letter is to present novel type of topological color
solitons existing in symplectic subgroup of the extended chiral group $E\chi 
$ , as well as in complete $E\chi $ . As a first step we consider isolated
solitons which are defined as solitons in vacuum background field. We
describe vacuum in a phenomenogical way \cite{Shifman+Vainshtein+Zakharov}  
through condensates and assume that cubic gluon condensate is zero.

The extended chiral group $E\chi $ is the group of all gauge and chiral
transformations leaving quark lagrangian invariant, if external fields are
transformed accordingly \cite{Novozhilov-95}. In absence of external fields, 
$E\chi $ is the group of global color and flavor transformations leaving
free quark lagrangian invariant. The basic Dirac spinor $\Psi $ is
8-component, as considered first by Pauli and Gursey : 
$\Psi={q  \choose \overline{q}^T}$, and the Dirac lagrangian $L$ can 
be rewritten as $L=\frac 12\Psi ^T \widehat{F}\Psi $ with $\widehat{F}$ 
represented in the block form $\widehat{F}=\left( 
\begin{array}{cc}
C\Phi & -\hat D^T \\ 
\hat D & \overline{\Phi }C
\end{array}
\right) $,where $\hat D$ $=i\gamma ^\mu \left( \partial _\mu +V_\mu +\gamma
_5a_\mu \right) $ is the Dirac operator with external fields, $"^T"$ means
transposition and $\Phi =\gamma ^\mu \left( \phi _{5\mu }+\gamma _5\phi _\mu
\right) $ contains external vector diquark fields, $\overline{\Phi }=\gamma
_0\Phi ^{+}\gamma _0$ .

To avoid difficulties with the Majorana spinors in finding a counterpart of $%
\widehat{F}$ in the Euclidean path integral over $\Psi $, one should take
special care $\left[ 7\right] $ keeping in mind that what is necessary for
chiral actions is only to calculate $\det \widehat{F}$ . To this end we use
the hermitean operator $\widehat{G}=\left( 
\begin{array}{cc}
\hat D & \overline{\Phi } \\ 
\Phi & \hat D_c
\end{array}
\right) ,\hat D_c=C^{-1}\hat D^TC,$ with $\det \widehat{G}=\det \widehat{F}$
and required positivity properties $\left[ 7\right] $. Thus, having in mind
the chiral anomaly and related effective action, one should study
transformations of operator $\widehat{G}$ induced by quark transformation 
$\delta \Psi $ 

\begin{eqnarray}
\delta \Psi =-\Omega \Psi ,\widehat{G}\rightarrow \widehat{G}^{\prime }=\exp
\left( -\Xi +\gamma _5\Theta \right) \widehat{G}\exp \left( \Xi +\gamma
_5\Theta \right) 
\end{eqnarray}
where antihermitean matrices $\Omega ,\Xi ,\Theta $ are given by $\Omega
=\rho _{11}\left( \alpha +\gamma _5\chi \right) +\rho _{22}\left( \alpha
^{*}-\gamma _5\chi ^{*}\right) +\rho _{12}\left( \xi +\gamma _5\omega
\right) C+\rho _{21}\left( -\xi ^{*}+\gamma _5\omega ^{*}\right) C$
$\Xi =\rho _{11}\alpha +\rho _{22}\alpha ^{*}+\rho _{12}\xi +\rho _{21}\xi
^{*},\Theta =\rho _{11}\chi -\rho _{22}\chi ^{*}+\rho _{12}\omega -\rho
_{21}\omega ^{*}$ . We introduced notation: $\rho _{ab}$ is a $2\times 2$
block matrix with elements$\left( \rho _{ab}\right) _{cd}=\delta _{ac}\delta
_{bd}$ . We have $\alpha ^{+}=-\alpha ,\chi ^{+}=-\chi ,\xi ^T=-\xi ,\omega
^T=\omega $ .The quark baryon number is $\hat b=\frac 13(\rho _{11}-\rho
_{22})=\frac 13\hat \rho _3$ .

Transformations with $\Xi $ do not change the quark path integral, while
transformations with $\Theta $ induce chiral anomalies. The Lie algebras
with $\Xi +\gamma _5\Theta $ and $\Xi +\Theta $ are isomorphic. For $N_C$
colors and $N_F$ flavors, generators $\Xi +\Theta $ are in algebra of $%
U\left( 2N\right) $ , $N=N_CN_F$ . Non-anomalous generators $\alpha $ are in
algebra of $SU\left( N\right) $ and include color $SU\left( N_C\right) $ ;
the left-right group ${\bf G}_{LR}=$ $SU\left( N\right) _L\times SU\left(
N\right) _R$ has $\alpha ,\chi $ as generators, while $\left( \Xi +\Theta
\right) _{sp}=\rho _{11}\alpha +\rho _{22}\alpha ^{*}+\rho _{12}\omega -\rho
_{21}\omega ^{*}$ is in algebra of symplectic group $Sp\left( 2N\right) $,
because $\left( \Xi +\Theta \right) _{sp}^Ti\rho _2+i\rho _2\left( \Xi
+\Theta \right) _{sp}=0$ . In $E\chi $ non-anomalous generators $\Xi $ are
in algebra of the orthogonal group $O\left( 2N\right) $ and include in
addition to block diagonal generators $\alpha $ also non-diagonal generators 
$\xi $ , which arise from non-commutativity of anomalous generators $\Theta
_{LR}=$ $\rho _{11}\chi -\rho _{22}\chi ^{*}$ and $\Theta _{sp}=\rho
_{12}\omega -\rho _{21}\omega ^{*}$  \cite{Novozhilov-95}. Anomalous
generators $\Theta _{sp}$ of the symplectic group $Sp\left( 2N\right) $
belong to the coset $Sp\left( 2N\right) /SU\left( N\right) $ . We see that
there are two distinguished subgroups ${\bf G}_{LR}$ and $Sp\left( 2n\right) 
$ of $E\chi $ with the same block diagonal non-anomalous generators $\Xi
^0=\rho _{11}\alpha +\rho _{22}\alpha ^{*}$ and different anomalous parts $%
\Theta _{LR}$ and $\Theta _{sp}$ , which do not require introduction of
additional non-anomalous generators $\xi $ . In the case of entire group $%
E\chi $ , anomalous generators $\Theta $ include both $\chi $ and $\omega $
and belong to the coset $SU\left( 2N\right) /O\left( 2N\right) $.

The chiral field is $U=\exp \Theta $ ; calculation of $\det \{\exp (\gamma
_5\Theta )\hat G\exp \left( \gamma _5\Theta \right) \}/\det \hat G$ by
integration of anomaly follows the standard bosonization procedure. After
eliminating external color axial fields, $a_\mu =0,$ we get the chiral
action for $2N$ internal degrees of freedom $W\left( U\right) =W_{eff}\left(
U\right) -W_{wz}$ with an effective lagrangian $L_{eff}\left(
U\right) $ and the Wess-Zumino term $W_{WZ}$

\begin{eqnarray}
L_{eff}\left( U\right) =tr_{C,F}\{\frac{f_\omega ^2}4D_\mu UD^\mu U^{-1} 
\nonumber   \\
+\frac 1{384\pi ^2}\left[ \frac 12\left[ UD_\nu U^{-1},UD_\mu U^{-1}\right]
^2-(UD_\nu U^{-1}UD^\nu U^{-1})^2\right] 
\nonumber   \\
+\frac 1{192\pi ^2}\left( [UD^\mu U^{-1},UD^\nu U^{-1}](\tilde G_{\nu \mu }+
U\tilde G_{\nu \mu }U^{-1})+\tilde G_{\mu \nu }U\tilde G^{\mu \nu
}U^{-1}\right) \} 
\end{eqnarray},

where coefficients contain additional factor $\frac 12$ coming from square
root of quark determinant ; $D_\mu =\partial _\mu +\left[ \tilde G_\mu
,*\right] ,\tilde G_\mu =\rho _{11}G_\mu +\rho _{22}\left( -G_\mu ^T\right)
+\rho _{12}\bar \Phi +\rho _{21}\Phi $ . The kinetic term depends on a
phenomenological parameter $f_\Theta ^2$ . As in the case of $G_{LR}$ color
solitons ...,we do not take into account the term which contains higher
derivatives and reflects presence of radial excitations.

Consider the symplectic group $Sp\left( 2N\right) $ with $\alpha $ in $%
SU\left( N\right) $, $N=N_CN_F$ . The chiral field $U_{sp}$ is a mapping
from $S^3$ to $\Theta _{sp}=\rho _{+}\omega -\rho _{-}\omega ^{*}$ with $%
\omega =\omega ^T$. For the whole group (with both $\alpha $ and $\omega $)
such mappings belong to $\pi _3={\bf Z}$ . Excluding $\alpha $ we get:
(a)Two or three colors, one flavor: $\omega =\lambda _a^{}\omega
_a,a=1,3,4,6,8$ , no spherical solitons .(b) Two colors, two flavors: $%
\omega =\Lambda _b\omega _b$ , $\Lambda _b$ are nine traceless symmetrical
matrices $\sigma _2\times \tau _2$ and $\sigma _{\tilde k}\times \tau _{%
\tilde l};\tilde k,\tilde l=0,1,3$ ; no spherical solitons .(c) Three
colors, two flavors; non-anomalous part $\Xi _{sp}$ is given by 12$\times 12$
matrix with $\alpha $ in $SU\left( 6\right) $ built on $SU\left( 3\right) $
matrices $\lambda _a$ and flavor Pauli matrices $\tau _k$, while anomalous
part $\Theta _{sp}$ together with $\Xi _{sp}$ span Sp$\left( 12\right) $
algebra. Symmetric matrix $\omega $ contains fields with diquark quantum
numbers associated with both symmetrical matrices $\lambda _a^S\times \tau
_k^S$ and both antisymmentrical ones $\lambda _a^A\times \tau _2$ .
Antisymmetric matrices $\lambda ^A$ $=\left( \lambda _2,-\lambda _5,\lambda
_7\right) \equiv \left( O_k\right) ;\left( O_k\right) _{ij}=-i\varepsilon
_{kij}$ in color $O\left( 3\right) $ algebra we combine with unit coordinate
vector $r_k$,$r_kr_k=1$ into $\hat r=O_kr_k$. We retain only those
parameters $\omega $ , which enter with generators of $O\left( 3\right) $ ,
introduce the shape function $F_{sp}\left( R\right) $ and write the
anomalous part $\Theta _{sp}$ and the chiral field as

\begin{eqnarray}
\Theta _{sp}=\rho _{+}i\tau _2O_k\vartheta _k-\rho _{-}i\tau _2O_k\vartheta
_k^{*}\equiv i\tau _2\eta \hat rF_{sp},\eta =\rho _1\cos \chi -\rho _2\sin
\chi 
\nonumber \\
U_{sp}=\exp \Theta _{sp}=1+i\tau _2\eta \hat r\sin F+\hat r^2\left( \cos
F-1\right) ,R^2=x_kx_k 
\end{eqnarray}

assuming that $\chi $ is constant. Isospin matrices $I_k=(\rho _3\tau
_1,\tau _2,\rho _3\tau _3)$ commute with $\Theta _{sp}$. $U_{sp}$ can
describe two conjugate color solitons $U_{\pm }=\frac 12\left( 1\pm \tau
_2\eta \right) \exp \left( \pm i\hat rF_{sp}(R)\right) $ made out of fields 
$\vartheta _k,\vartheta _k^{*}$ with diquark quantum numbers. 
$\Theta _{sp}$
satisfies $\Theta _{sp}^3=-i\Theta _{sp}F_{sp}^2$ . In this special case
transformations close in a smaller group, namely $O\left( 6\right) \sim
SU\left( 4\right) $ , and $\alpha $ belong to $U(3)\sim SU(3)\times U\left(
1\right) $ ,while Goldstone diquark degrees of freedom 
$\eta \Theta _{sp}$
belong to the complex projective space $CP^3=SU\left( 4\right) /SU\left(
3\right) \times U\left( 1\right) $ \cite{Novozhilov-95}. Note that $\eta $ anticommutes
with the quark baryon number $\hat b=\frac 13\rho _3$ . Thus, the commutator 
$\left[ \hat G_k,U_{sp}\right] $ with the background vacuum field $\hat G%
_k=\rho _{11}G_k+\rho _{22}\left( -G_k^T\right) ,G_k^a=V_k\hat N^a,\hat N
=\lambda _aN_a,N_aN_a=1$ contains anticommutator $\{N+N^T,\hat r\}$ in order
to preserve the form of $U_{sp}$ we have to restrict $\Xi _{sp}$ to a common
subgroup $O\left( 3\right) $ of $SU\left( 3\right) \times SU\left( 3\right)
^{*}$ . This is possible for vacuum background field $G_k^a=V_k\hat N^a,\hat 
N=\lambda _aN_a,tr\hat N^3=0$ in the gauge $G_0^a=0.$ Then $\hat G_k=\rho
_{11}G_k+\rho _{22}\left( -G_k^T\right) =\left( \rho _{11}+\rho _{22}\right)
V_k\tilde N,$ where $\tilde N=O_lN_l,\tilde N^3=\tilde N$ . Finally, the
soliton subgroup ${\bf G}_{sol}$ in symplectic $Sp\left( 12\right) $ is
given by $(\Xi +\gamma _5\Theta )_{sp}^{sol}=$ $iO_k\left( \alpha _k+\gamma
_5\tau _2\eta r_kF_{sp}\right) +i\rho _3\tau _3\alpha _F$ , where $\alpha _F$
is a flavor parameter. For comparison, in the left-right group $(\Xi +\gamma
_5\Theta )_{LR}$ with both $\alpha $ and $\chi $ in $SU\left( 3\right) $
algebra the analogue soliton subgroup is $\left( \Xi +\gamma _5\Theta
\right) _{LR}^{sol}=iO_k\left( \alpha _k+\gamma _5\rho _3r_kF_{LR}\right) $
.Thus, ${\bf G}_{sol}=O\left( 3\right) _L\times O\left( 3\right) _R\times
U\left( 1\right) _F$ , but in the left-right group ${\bf G}_{sol}$ the
baryon direction $\rho _3$ is replaced by another one $\tau _2\eta $ . If we
introduce the baryon $SU\left( 2\right) _b$ with generators $(\tau _2\rho
_1,\tau _2\rho _2,\rho _3)/2i$ , then $\tau _2\eta $ can be reached from 
$\rho _3$ by rotation.

Complete $E\chi $ group, anomalous generators $\Theta =\left( \chi ,\omega
\right) $ . Two or three colors, one flavor: $\Theta $ is built on
generators $\left( \rho _3\lambda ^A,\lambda ^S,\rho _{\pm }\lambda
^S\right) $, no spherical solitons, no constituent quarks, if generators
with $\rho _{\pm }$ are present and $E\chi $ is not reduced to the
left-right subgroup. Two colors, two flavors: no SU(2) and O(3) subgroups in
the same sense. Three colors, two flavors : most general $\Theta $ is built
on generators $\left( \rho _3\lambda ^A\tau ^S,\rho _3\lambda ^S\tau ^A,\rho
_{\pm }\lambda ^S\tau ^S,\rho _{\pm }\lambda ^A\tau ^A\right) $ and among
them there are two O(3) subgroups with generators $\Theta _{\pm }\left(
O\left( 3\right) \right) \sim \frac 12\left( 1\pm \zeta \right) \left(
\lambda _2,-\lambda _5,\lambda _7\right) ,$ where $\zeta =\rho _31_\tau \cos
\kappa +\eta \tau _2\sin \kappa $ , $\left[ \Xi ,\zeta \right] =0$ , and $%
\eta =\rho _1\cos \chi -\rho _2\sin \chi ,\zeta ^2=1$ , angles $\kappa $ and 
$\chi $ are two constant parameters; $\chi $ comes from the symplectic group
Sp(12), while $\kappa $ describes mixing of left-right and symplectic
contributions to $\zeta $ : when $\kappa =0$ , no symplectic contribution is
present: when $\kappa =\pi /2$ , there is no left-right contribution. The
chiral field for $E\chi $ in this case is $U=\exp i\zeta \hat rF\left(
R\right) $ , it can be obtained by $SU\left( 2\right) _b$ rotation from $%
U_{sp}$ or $U_{LR}$. In this special case $\Xi +\gamma _5\Theta =iO_k\left(
\alpha _k+\gamma _5\zeta r_kF\right) $, when $\Xi $ is invariant under
rotations of baryon $SU\left( 2\right) _b$ , no additional gauge parameters $%
\xi $ arise from repetition of such transformation. Thus, although we
started from different subgroups of the extended chiral group for $%
N_C=3,N_F=2$, the resulting structure of solitonic chiral subgroups is the
same: it is $O\left( 3\right) _{\tilde L}\times O\left( 3\right) _{\tilde R}$
, where left $\tilde L$ and right $\tilde R$ are defined by projectors $%
\frac 12\left( 1\pm \gamma _5\tilde \rho \right) $ , where a block matrix $%
\tilde \rho $ with $\tilde \rho ^2=1$ , is linear in the Pauli matrices $%
\rho _1\tau _2,\rho _2\tau _2,\rho _3$ of the baryon $SU\left( 2\right) _b$
. The reason is that $E\chi $ includes quarks and antiquarks.

Topological charge $t_\chi \left( U\right) $ for a soliton $U$ in the
left-right subgroup of $E\chi $ is related to the quark baryon number $%
b=\rho _3/N_C$ 

\begin{eqnarray}
t_\chi \left( U\right) =\frac 12\frac 1{24\pi ^2N_C}\int d^3x\varepsilon
_{ijk}tr\{\rho _3UD_iU^{+}UD_jU^{+}UD_kU^{+}\} 
\end{eqnarray}

and coincides with the baryon number of soliton $U$ . For three colors and $%
N_F$ flavors $t_\chi $ starts from $\frac 23N_F$ (flavor analogous case is
skyrmion as dibaryon studied by Balachandran et al.[8 ]) . The topological
charges for solitons in symplectic subgroup $Sp\left( 12\right) $ and
complete group $E\chi $ are winding numbers in directions $\rho _{\perp
}=\tau _2\eta $ and $\rho _E=\zeta $ within the baryon $SU\left( 2\right) _b$
-space 

\begin{eqnarray}
t_A\left( U\right) =\frac 12\frac 1{24\pi ^2N_C}\int d^3x\varepsilon
_{ijk}tr\{\rho _AU\partial _iU^{+}U\partial _jU^{+}U\partial
_kU^{+}\}, && A=\perp ,E. 
\end{eqnarray}

and do not describe solitonic baryon numbers. These winding numbers start
from $\left| \frac 43\right| $ . Boundary conditions for the shape functions 
$F\left( R\right) $ in both cases trivally follow from \cite{Balachandran-83+85}.

Finiteness of soliton mass follows from the static effective lagrangian $%
L_{eff}$ and asymptotic behavior of the shape function $F\left( R\right) $
at large $R.$ The mass is given by the positive definite functional, as it
can be easily verified; asymptotics of $F$ is defined by the kinetic term.
The kinetic term averaged over directions $N^k$ and $\nu _t$ of background
vacuum field $G_l^k=V_lN^k,V_l=-\frac \pi {2i}\varepsilon _{ljt}r_j\nu _tR%
\sqrt{\frac{C_g}2}$ in $O\left( 3\right) $ color and coordinate spaces takes
the following form 
\begin{eqnarray}
\ \overline{K} &=&\frac 14N_Ff_\omega ^2\{2F^{\prime ^2}+2\left( \frac 2{R^2}%
+\frac 19\pi ^2C_gR^2\right) \sin ^2F+ 
 +(\frac 4{R^2}+\frac 19\pi ^2C_gR^2)\left( \cos F-1\right) ^2\}
\end{eqnarray}
where $C_g$ is the gluon condensate. Thus, asymptotic behavior at large $R$
of the shape function $F\left( R\right) $ is governed by the same equation
as in the case of color $SU\left( 2\right) $ soliton in  \cite{Novozhilov-01}.
 We use the result 
\begin{eqnarray}
F\rightarrow \left( f_0R\right) ^{-\frac 32}\exp \left( -\frac \pi 3\sqrt{%
\frac{C_g}2}R^2\right) ,R\rightarrow \infty 
\end{eqnarray}
which guarantees that the mass $M=-4\pi \int dRR^2L_{stat}$ is finite for
positive condensate $C_g$, i.e. for chromomagnetic vacuum field. It follows
from the effective lagrangian that the soliton mass is invariant under 
$SU\left( 2\right) _b$rotation.

We have shown that classically stable, finite mass topological solitons
exist in the extended chiral group $E\chi $ . In the case of three colors,
two flavors their status is described by the special chiral group $O\left(
3\right) _{\tilde L}\times O\left( 3\right) _{\tilde R}$ with notions of
left $\tilde L$ and right $\tilde R$ defined in terms of $\frac 12(1\pm $ $%
\gamma _5\zeta )$ , where $\zeta $ is a constant non-abelian charge
direction in the $SU\left( 2\right) _b$ and $\rho _3$ is baryon number
direction. In the case of isolated soliton, $\zeta =$const. will be a
covariant notion, when $\zeta $ commutes with direction $\hat N$ of
background vacuum $SU\left( 3\right) $ field, as it is for tr$\hat N^3=0.$
The vacuum background field defines also an asymptotic behavior of the shape
function $F$ ; the field should be chromomagnetic. This pattern based on
establishing mapping from $S^3$ to anomalous part $\Theta $ of $E\chi $ and
using properties of vacuum background field can be followed in more
complicated cases. This expansion of the world of color solitons is not
accompanied by the widening of pure QCD gauge space: it is still $SU(3)$ ;
no additional gauge degrees $\xi $ were still required to accomodate novel
solitons. However, these degrees $\xi $ are likely to appear, if vacuum is
to be described by set of condensates. Novel type of solitons may be essential 
in discussion of baryon assymetry and baryon number nonconservation.

We thank M.Bordag for the support and hospitality extended to one of the authors 
(YuN) during his visit to University of Leipzig. We thank M.Bordag and D.Vassilevich
for interesting discussions.


\begin{references}
\bibitem{Adler+Bell+Jackiw}  S.L.Adler, Phys.Rev. {\bf 177}, 2426 (1969);
J.S.Bell and R.Jackiw, Nuovo Cim., {\bf A60}, 47 (1969)

\bibitem{Faddeev+Niemi+Battye+Sutcliffe}  L. Faddeev and A. Niemi, Nature 
{\bf 387}, 58 (1997); R.Battye and P.Sutcliffe, Phys.Rev.Lett.{\bf 81, }4798
(1998); L. Faddeev and A. Niemi, Phys.Rev.Lett. {\bf 85, }3416 (2000);
hep-th/0101078

\bibitem{Cho}  Y.M. Cho, Phys.Rev.Lett. {\bf 87}, 252001 (2001); W.S.Bae,
Y.M.Cho, and S.W.Kimm, Phys.Rev. {\bf D65, }025005 (2002); Y.M.Cho,H.W.Lee
and D.G.Pak, Phys.Lett. {\bf B525}, 347 (2002); hep-th/0201179.

\bibitem{Kaplan}  D.B.Kaplan, Phys.Lett. {\bf B235}, 163 (1990); Nucl.Phys. 
{\bf B351}, 357 (1991); Y.Frishman, A.Hanany and M.Karliner, hep-ph/9507206.

\bibitem{Novozhilov-01}  V.Novozhilov and Yu.Novozhilov, Phys.Lett. 
{\bf B522}, 49 (2001), hep-ph/0110006;  Teor.Math.Phys. {\bf 131}, 62 (2002).

\bibitem{Shifman+Vainshtein+Zakharov}  M.A. Shifman, A.I.Vainshtein and
V.I.Zakharov, Nucl.Phys. {\bf B147},385 (1979)

\bibitem{Novozhilov-95}  Yu.Novozhilov, A. Pronko and D.Vassilevich,
Phys.Lett. {\bf B343}, 358 (1995);(E) {\bf B351}, 601 (1995)

\bibitem{Ball+NPV}  R.Ball, Phys.Lett. {\bf B227} , 445 (1989); Yu.
Novozhilov, A. Pronko and D. Vassilevich, Phys.Lett. {\bf B321}, 425 (1994)

\bibitem{Balachandran-83+85}  A.P.Balachandran, V.P.Nair, S.G.Rajeev and
A.Stern, Phys.Rev. {\bf D27},1153 (1983); A.P.Balachandran, F.Lizzi,
V.G.J.Rodgers and A.Stern, Nucl.Phys. {\bf B256}, 525 (1985)
\end{references}
\end{document}